\documentclass[a4paper,11pt]{article}
\usepackage{pos}
\usepackage[version=4]{mhchem}
\usepackage{multirow}
\usepackage{stackengine}

\title{Correlating lepton flavor violating
$b \to s$ and leptonic decay modes
in a minimal abelian extension of the Standard Model}

\ShortTitle{\footnotesize{Correlating lepton flavor violating $b\to s$ and leptonic decay ... abelian extension of the SM}}

\author*[a,b]{Davide Milillo}

\affiliation[a]{Istituto Nazionale di Fisica Nucleare, Sezione di Bari, Via Orabona 4, 70126, Bari, Italy}

\affiliation[b]{Dipartimento Interuniversitario di Fisica ``Michelangelo Merlin'',\\ Università degli Studi di Bari ``Aldo Moro''
Via Orabona 4, 70126, Bari, Italy}

\emailAdd{davide.milillo@ba.infn.it}

\abstract{We examine possible correlations between $b \to s \ell_1^- \ell_2^+$ transitions -- both in the lepton flavor conserving ($\ell_1=\ell_2$) and violating case ($\ell_1 \neq \ell_2$) -- and purely leptonic flavor violating decays within the ABCD model \cite{ABCD}, a minimal abelian extension of the Standard Model (SM) introducing a new $\text{U}(1)'$ symmetry. The associated neutral $Z'$ boson has generation-dependent, flavor non-universal couplings to SM fermions, governed by three rational parameters $\epsilon_{1,2,3}$, which sum to zero to ensure gauge anomaly cancellation. Each $\epsilon_i$ is common to all fermions of a given generation, thus inducing correlations among quark and lepton observables. For lepton flavor conserving (LFC) processes, only small deviations from SM predictions were found \cite{DFP}, reflecting the mutual constraints between the quark and lepton sectors, which preclude large discrepancies. On the other hand, the model allows tree-level lepton flavor violating (LFV) decays, yielding correlations between LFV $b\to s$ transitions and charged lepton decays. The analysis of such correlations shows that the current experimental upper bounds for the rates of $\tau^- \to \mu^-\mu^+\mu^-$, \  $\mu^-\to e^- \gamma$, \ $\mu^- \to e^- e^+e^-$ and $\mu^- \to e^-$ conversion in nuclei constrain branching ratios of LFV $B_{(s)}$ decays in hierarchical order \cite{DFP}.}

\FullConference{Fourth Italian Workshop on Physics at High Intensity (WIFAI2025)\\
11–14 November 2025\\
Bari, Italy\\}

\begin{document}
\maketitle


\section{Introduction}
The Standard Model (SM) is the most accurate theory describing all known fundamental particles and their interactions (except for gravity). However, despite remarkable experimental success, it still leaves many questions unanswered. For instance, the SM does not explain the number of fermion generations and the hierarchy of their masses. Furthermore, although the SM incorporates $CP$ violation through the complex phase of the Cabibbo-Kobayashi-Maskawa (CKM) matrix, the degree of violation is too small to account for baryogenesis. The hierarchical structure of
CKM is unexplained too. A new theory capable of addressing, at least partially, such open problems
is actively sought for, the SM being its low-energy limit. Many tensions have emerged between SM predictions and experimental data, especially in the flavor sector. Flavor physics thus represents a promising path for probing New Physics (NP): processes such as flavor changing neutral current (FCNC) transitions, which are loop-suppressed in the SM, are extremely sensitive NP effects, as new particles may contribute as virtual states \cite{Buras:2020book}. The list of flavor anomalies includes \cite{Capdevila:2023rev, Colangelo:2024ped}:\vspace{-4pt}
\begin{itemize}
    \item Violation of the CKM unitarity (Cabibbo anomaly) and puzzles in the determination of its parameters, e.g. the long-standing tension between the inclusive and exclusive determinations of $|V_{ub}|$ and $|V_{cb}|$; \vspace{-4pt}
    \item Discrepancies in FCNC $b \to s$ transitions, e.g. suppressed decay rates of $B\to K \mu^+ \mu^-$ and $B_s\to \phi \mu^+ \mu^-$ alongside deviations in $B\to K^* \mu^+ \mu^-$ angular observables \cite{HFLAV:2024ctg}; \vspace{-4pt}
    \item Hints of lepton flavor universality (LFU) violation in $b \to c \ell \bar \nu_\ell$ induced decays, with the ratios $R_{D^{(*)}}$ exceeding the SM prediction at the level of $3\sigma$ \cite{HFLAV:2024ctg}.
\end{itemize}

Although none of these tensions, taken individually, constitutes definitive evidence for NP, their number and persistence strongly motivate the search for physics beyond the SM (BSM). In this context, rare $K$ and $B$ meson decays represent a highly promising avenue. While several observables involved in these transitions are extremely sensitive to NP, they depend on SM parameters and non-perturbative QCD quantities, requiring careful control of the associated uncertainties \cite{Buras:2021app, Buras:2022wpw}.

Among the possible extensions of the SM, we consider the ABCD model \cite{ABCD}, a minimal
abelian extension of the SM predicting a new $Z'$ gauge boson with generation-dependent, flavor
non-universal couplings to quarks and leptons. These specific couplings induce correlations between quark and lepton sectors, preventing large deviations from SM predictions. Nevertheless, the model still provides a clear window into NP through tree-level LFV transitions, which are strictly forbidden in the SM. Here we summarize the main findings in Ref. \cite{DFP}, focusing on the correlations between lepton flavor conserving (LFC) and violating $B_{(s)}$ decays, as well as between LFV $b \to s$ transitions\footnote{\ Recent studies of lepton flavor violating $b \to s$ modes can be found in Refs. \cite{LFV2, LFV3, LFV4, LFV5, LFV6, LFV7}} and purely leptonic decays.


\section{The ABCD model}
In the ABCD model, the SM gauge group is extended by an extra $\text{U}(1)'$ symmetry, resulting in a new neutral gauge boson $Z'$ with gauge coupling $g_{Z'}$ and mass $M_{Z'}$. Many models introduce a new $Z'$ boson, and they typically differ in the assignment of the fermion charges under $\text{U}(1)'$ -- the $z$-hypercharges \cite{Zp1,Zp2,Zp3}. In ABCD, $z$-hypercharges are chosen to cancel gauge anomalies through the rational solution of six anomaly cancellation equations (ACEs). For each fermion generation, the $z$-hypercharge is given by the sum of the SM weak hypercharge and a generation-dependent parameter $\epsilon_i$ ($i=1,2,3$), constrained to fulfill the anomaly cancellation condition $\epsilon_1+\epsilon_2 + \epsilon_3 =0$.

Besides SM fermions, the ABCD model comprises three heavy right-handed neutrinos. After rotating from flavor to mass eigenstates, the $Z'$ couplings to fermions become flavor non-universal, generating FCNC and LFV transitions at tree level. Specifically, couplings to left-handed fermions depend on the CKM and Pontecorvo-Maki-Nakagawa-Sakata (PMNS) matrix elements, while those to right-handed fermions involve additional mixing angles and phases, namely $(\tilde{s}_{ij}, \phi_{K,d,s})$ for quarks and $(\tilde{t}_{ij}, \phi_{ij})$ for leptons.

Overall, the new parameters of the model are $\left\{ g_{Z'}, \ M_{Z'}, \ \epsilon_1, \ \epsilon_2 \right\}$, alongside CKM/PMNS elements and the new mixing parameters for right-handed fermions. Among the different coupling configurations in Ref.\cite{ABCD}, we focus on Scenario A, where no flavor violation is allowed for right-handed fermions. Nevertheless, correlations between quark and lepton observables emerge.


\section{Observables}
In the ABCD model, the new neutral $Z'$ boson mediates FCNC and LFV transitions at tree level. Our analysis focuses on $b\to s \ell_1^+ \ell_2^-$ induced decays, namely $B_s\to \ell^+_1 \ell^-_2$ and $\bar{B}\to \bar{K}^* \ell^+_1 \ell^-_2$, both in the LFC and LFV cases. The low-energy Hamiltonian governing such modes comprises current-current, penguin, magnetic, and semileptonic operators:
\begin{equation}
    \mathcal{H}_{\text{eff}}=
    -\frac{4G_F}{\sqrt{2}} V_{tb}V_{ts}^*
    \sum_{i}[C_i \mathcal{O}_i+
    C_i' \mathcal{O}_i'] + \text{h.c.},
    \label{Heff}
\end{equation}
where $G_F$ is the Fermi constant and $V_{tq}$ ($q=b,s$) denotes the CKM matrix element (doubly Cabibbo-suppressed terms proportional to $V_{ub}V_{us}^*$ are neglected) \cite{Ball09}. In the SM, the dominant operators in Eq.(\ref{Heff}) are $\mathcal{O}_{7,9,10}$, and the corresponding Wilson coefficients are flavor-universal. Beyond the SM, primed operators $\mathcal{O}_{i}'$ of opposite chirality to $\mathcal{O}_i$, generally appear, but in Scenario A of the ABCD model their contribution vanishes. Therefore, NP contributions are entirely captured by a flavor-dependent shift in $C_{9,10}$ (as the NP contribution to $C_{7}$ is negligible).

For the mode $B_s \to \ell^+_1 \ell^-_2$, the SM decay rate (with $\ell_1 = \ell_2$) is solely governed by the $\mathcal{O}_{10}$ operator. Conversely, in the ABCD model, where LFV transitions are allowed, a non-vanishing contribution from $\mathcal{O}_{9}$ emerges. For $\bar{B} \to \bar{K}^* \ell^+_1 \ell^-_2$, several observables can be defined that are sensitive to NP, namely the lepton forward-backward asymmetry $A_{FB}$, the $K^*$ meson polarization fractions, and the $P$-observables \cite{Matias:2012xw}. These observables depend on $C_{7,9,10}$, allowing for simultaneous constraints on multiple Wilson coefficients. 

Regarding LFV charged lepton decays, the SM extension with massive neutrinos allows these transitions only at the 1-loop level. In the ABCD model, the radiative decay $\mu\to e\gamma$ similarly proceeds at the 1-loop level via a $Z'$ penguin diagram. Conversely, the purely leptonic decays $\tau \to 3\mu$ and $\mu \to 3e$ occur at tree-level through $Z'$ boson exchange. Finally, $\mu\to e$ conversion in nuclei is mediated at tree level by a $Z'$ boson and is sensitive to both lepton and quark couplings. Detailed discussions of leptonic observables can be found in Refs. \cite{ABCD, DFP}.


\section{Numerical analysis and selected results}
We outline the strategy for numerical analysis and present the results for selected observables. The model parameters are $\left\{ g_{Z'}, \ M_{Z'}, \ \epsilon_1, \ \epsilon_2 \right\}$, together with the CKM and PMNS matrix elements. For the PMNS matrix, we adopt the global fit results in Ref. \cite{Esteban:2024nu}. Regarding the CKM matrix, we fix the Wolfenstein parameter $\lambda=0.2253$ and the phase $\gamma=64.4^\circ$, while $|V_{cb}|$, $|V_{ub}|$ are varied within the range spanned by their exclusive and inclusive determinations \cite{HFLAV:2024ctg}:
\begin{equation}
    |V_{cb}|\in \bigl[ |V_{cb}|_{\text{exc}},
    |V_{cb}|_{\text{inc}}\bigr], \quad
    |V_{ub}|\in \bigl[ |V_{ub}|_{\text{exc}},
    |V_{ub}|_{\text{inc}}\bigr].
\end{equation}
Furthermore, we vary $g_{Z'} \in [0.01,0.1]$ and consider two benchmark masses $M_{Z'}=1$ and $3$ TeV. The $(\epsilon_1, \epsilon_2)$ parameter space is constrained by experimental data on $\Delta F=2$ observables, namely:
\begin{itemize}
    \item The mass differences $\Delta M_{d}$, $\Delta M_{s}$, and $\Delta M_{K}$ in the neutral $B_d$, $B_s$, and $K$ meson systems;\vspace{-6pt}
    \item The $CP$ asymmetries $S_{J/\psi K_s}$ and $S_{\psi \phi}$;\vspace{-6pt}
    \item The $CP$ violation parameter in the kaon sector $\varepsilon_K$.
\end{itemize}
These observables govern the neutral meson $M$-$\bar{M}$ mixing ($M=B_{d},B_{s},K$). While in the SM this mixing proceeds exclusively at the 1-loop level via box diagrams \cite{Buras:2020book}, the ABCD model introduces an additional tree-level contribution mediated by $Z'$ exchange. In our analysis, we require $\Delta M_{d,s}$, $S_{J/\psi K_{s}}$, and $S_{\psi \phi}$ to lie within a $5\%$ range of their experimental central values, $\Delta M_K$ within a $25\%$ range of the SM value (accounting for long-distance effects), and $\varepsilon_K \in [2.0, 2.5] \times 10^{-3}$ \cite{PDG}. These requirements determine the allowed regions in the $(\epsilon_1, \epsilon_2)$ plane.

\begin{figure}[h]
    \centering
    \includegraphics[width=0.32\linewidth]{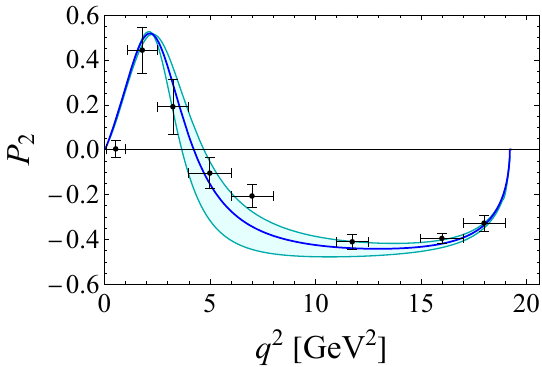}
    \includegraphics[width=0.32\linewidth]{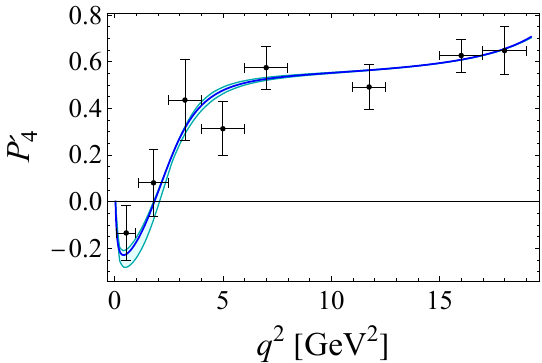}
    \includegraphics[width=0.32\linewidth]{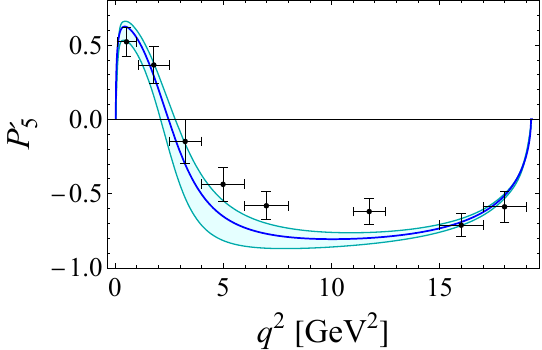}
    \caption{Predictions for the $P$-observables in $\bar{B} \to \bar{K}^*(K\pi) \mu^+ \mu^-$. The blue curves represent the SM central value, while the cyan band represents NP deviations in the ABCD model for $M_{Z'}=1$ TeV. The experimental data points correspond to the combined Run 1 and 2016 measurements from the LHCb collaboration \cite{LHCb:2020ao}.}
    \label{fig:LHCb2020}
\end{figure}

Once $\Delta F=2$ constraints are imposed, the NP contributions to the real part of the coefficients $C_{9,10}$ in the lepton flavor-conserving (LFC) case, $\text{Re}(C_{9,10}^{\text{\footnotesize{NP}}})_{\ell \ell}$ ($\ell=\mu,\tau$), are found to be modest -- roughly at the $10\%$ level of the corresponding SM coefficients. This highlights a core feature of the ABCD model: as pointed out in Ref. \cite{ABCD}, quark and lepton sectors mutually constrain each other, precluding large deviations from the SM predictions. This is consistent with the small scale of the discrepancies suggested by current flavor anomalies. As an example, in Figure \ref{fig:LHCb2020} we show the results for some of the $P$-observables in the $\bar{B} \to \bar{K}^*(K\pi) \mu^+ \mu^-$ decay, together with the measurements provided by the LHCb collaboration \cite{LHCb:2020ao}. The form factors parameterizing the $B\to K^*$ hadronic matrix elements are those in Ref. \cite{BSZ}; the values of other input quantities are quoted in Ref. \cite{DFP}.

\begin{figure}[ht]
    \centering
    \includegraphics[width=0.4\linewidth]{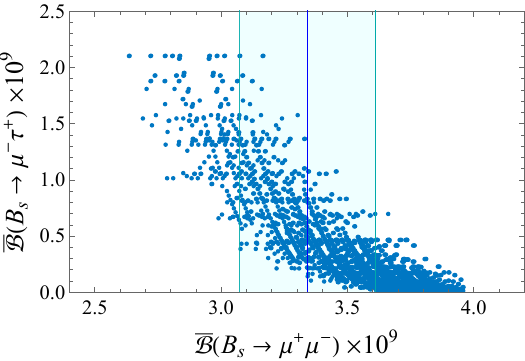}
    \qquad
    \includegraphics[width=0.4\linewidth]{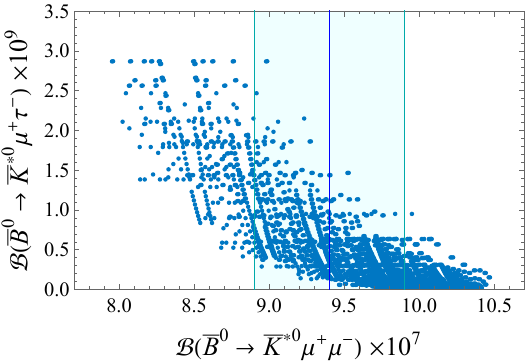}
    \caption{Correlations between the branching ratios of (left) $B_{s} \to \mu^- \mu^+$ and $B_{s} \to \mu^- \tau^+$ and of (right) $\bar{B} \to \bar{K}^* \mu^- \mu^+$ and $\bar{B} \to \bar{K}^* \mu^- \tau^+$ for $M_{Z'}=1$ TeV obtained after imposing $\Delta F=2$ constraints. The shaded vertical band corresponds to the experimental world averages \cite{PDG}.}
    \label{fig:Bcorr}
\end{figure}

Furthermore, we investigate the correlations between LFC $B_{(s)}$ decays and their LFV counterparts, finding that the LFV modes are effectively bounded by the LFC ones. In Figure \ref{fig:Bcorr} we illustrate the correlations between the LFC and LFV branching ratios for $B_{s} \to \ell_1^+\ell_2^-$ and $\bar{B} \to \bar{K}^*\ell_1^+\ell_2^-$.

\begin{figure}[t]
    \centering
    \includegraphics[width=0.31\linewidth]{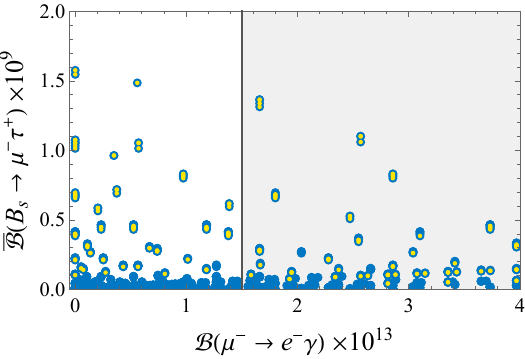} \quad
    \includegraphics[width=0.31\linewidth]{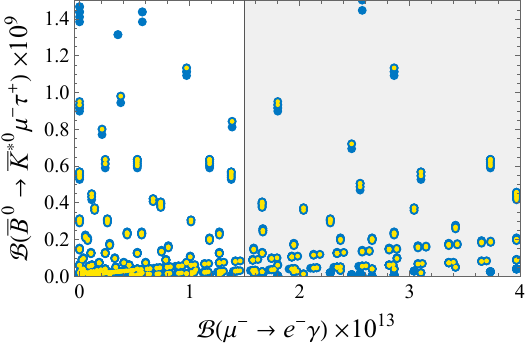}\\
    \includegraphics[width=0.31\linewidth]{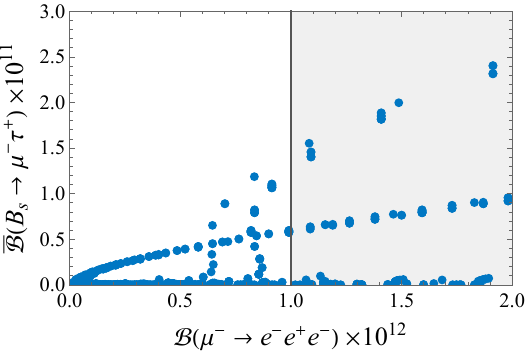} \quad
    \includegraphics[width=0.31\linewidth]{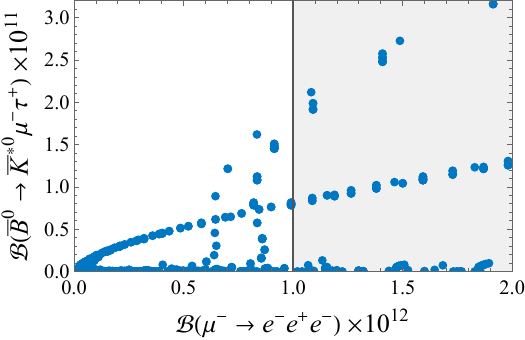}\\
    \includegraphics[width=0.31\linewidth]{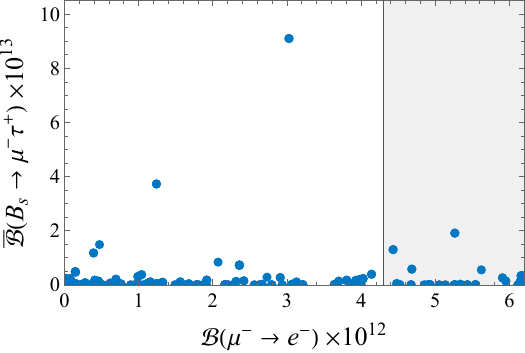} \quad
    \includegraphics[width=0.31\linewidth]{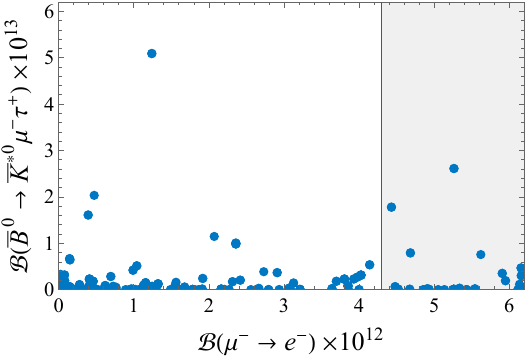}    
    \caption{Correlations between branching ratios of LFV $B_{(s)}$ and leptonic decays. The shaded gray band represents the experimental upper bound quoted in Table \ref{Tab:BRBlept}. The yellow points in the upper panels are selected from the blue ones by requiring $\bar{\mathcal{B}}(B_{s} \to \mu^+\mu^-)$ and $\mathcal{B}(\bar{B} \to \bar{K}^*\mu^+\mu^-)$ to agree with experimental measurements within $1\sigma$. We omit the plot for $\tau \to 3\mu$, as it does not impose any significant constraint.}
    \label{fig:Bleptcorr}
\end{figure}

\begin{figure}
    \centering
    \includegraphics[width=0.4\linewidth]{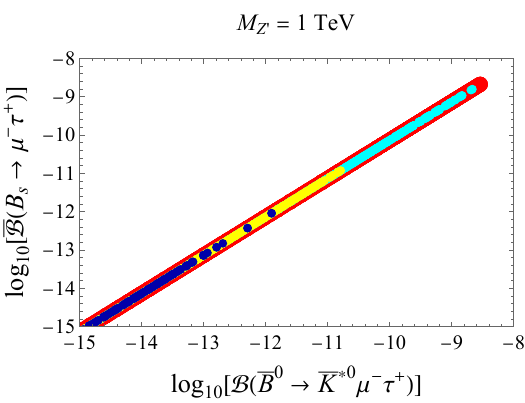}
    \quad
    \includegraphics[width=0.51\linewidth]{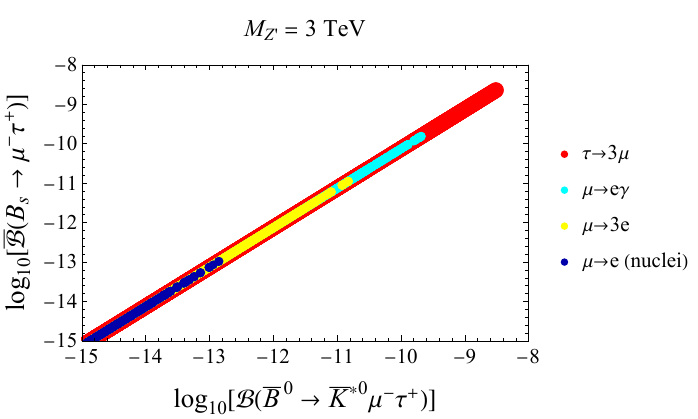}
    \caption{Correlations between branching ratios of LFV $B_{(s)}$ decays when the constraints from LFV lepton decays are successively imposed. The results are shown for $M_{Z'}=1$ TeV (left panel) and 3 TeV (right panel).}
    \label{fig:logBleptcorr}
\end{figure}

Finally, we examine the correlations between LFV $b\to s$ transitions and the purely leptonic decay modes. We find that the leptonic modes play an increasingly important role in constraining the $B_{(s)}$ branching ratios. Specifically, while the experimental upper bounds on $\tau \to 3 \mu$ \cite{PDG} and $\mu \to e \gamma$ \cite{MEG:2025meg} still permit LFV $B_{(s)}$ branching ratios of $O(10^{-9})$, significantly more stringent constraints stem from $\mu \to 3e$ \cite{SINDRUM:1988m3e} and $\mu \to e$ conversion in Ti \cite{SINDRUM:1998me}, as they push $B_{(s)}$ branching ratios down to $O(10^{-11})$ and $O(10^{-13})$, respectively. The results are shown in Figure \ref{fig:Bleptcorr} and detailed in Table \ref{Tab:BRBlept}. A visual summary of the hierarchical role of LFV lepton decays is shown in Figure \ref{fig:logBleptcorr}.

\begin{table}[h]
    \centering
    \begin{tabular}{c|c|c|c|c|c}
        \hline
         LFV mode &  $\mathcal{B}^{(1)} \times 10^9$ & $\mathcal{B}^{(2)} \times 10^9$ & $\mathcal{B}^{(3)} \times 10^{11}$ & $\mathcal{B}^{(4)} \times 10^{13}$ & experiment \cite{PDG}\\[2pt]
         \hline
         
         $B_s \to \mu^- \tau^+$ & $0.00 \div2.10$ & $0.00\div1.60$ & $0.00\div1.20$ & $0.00\div9.20$ & $<4.2\times 10^{-5}$\\

         $\bar{B} \to \bar{K}^*\mu^- \tau^+$ & $0.00 \div2.90$ & $0.00\div1.15$ & $0.00\div1.60$ & $0.00\div5.10$ & $<1.0\times 10^{-5}$\\[2pt]
         \hline

         $B_s \to \mu^- \tau^+$ & $0.00 \div2.30$ & $0.00\div1.14$ & $0.00\div1.10$ & $0.00\div1.05$ & $<4.2\times 10^{-5}$\\

         $\bar{B} \to \bar{K}^*\mu^- \tau^+$ & $0.00 \div3.10$ & $0.00\div0.20$ & $0.00\div1.53$ & $0.00\div1.42$ & $<1.0\times 10^{-5}$\\[2pt]
         \hline
    \end{tabular}
    \caption{Ranges of LFV $B_{(s)}$ decay rates in Scenario A of the ABCD model for $M_{Z'}=1$ TeV (first two rows) and $M_{Z'}=3$ TeV (last two rows) \cite{DFP}. Ranges are obtained by imposing constraints from experimental upper bounds on LFV lepton decays: (1) from $\tau \to 3\mu$ \cite{PDG}, (2) from $\mu \to e \gamma$ \cite{MEG:2025meg}, (3) from $\mu \to 3e$ \cite{PDG, SINDRUM:1988m3e}, and (4) from $\mu \to e$ conversion in Ti \cite{SINDRUM:1998me}. The last column shows the $90\%$ C.L. experimental limit \cite{PDG}. \vspace{-8pt}}
    \label{Tab:BRBlept}
\end{table}

\section{Conclusions}
We have presented selected results obtained in Ref. \cite{DFP} within the ABCD model \cite{ABCD}, a minimal abelian extension of the SM, which predicts a new gauge boson $Z'$ with generation-dependent, flavor-conserving and violating couplings to quarks and leptons. The particular solution adopted for ACEs implies the existence of correlations between quark and lepton sectors. As a result, the two sectors constrain each other, preventing large deviations from SM -- a result consistent with the small discrepancies observed in Flavor Physics. The model, however, still leaves room for NP, as it predicts LFV transitions at the tree level. Correlations emerge, for example, LFV $B_{(s)}$ and charged lepton decays. Such correlations provide important information to constrain physical observables targeted by current and future experiments.

\section*{Acknowledgements}
D.M. thanks the coauthors of Ref. \cite{DFP}, and A.J. Buras for insightful discussions and valuable advices. This study has been carried out within the INFN project (Iniziativa Specifica) SPIF.


\end{document}